# Structure de l'Univers : quand l'observation guide la théorie... ou pas

Yaël Nazé (chercheur qualifié FNRS)
Université de Liège, Département AGO, Allée du 6 Août 17 Bat B5C, 4000-Liège, Belgique

1. Résumé/Introduction

Aux enfants, on présente parfois le processus scientifique comme un travail linéaire, commençant par une question et s'achevant avec l'élaboration d'une théorie, en passant par quelques expériences. La réalité de la construction de la science s'avère bien plus complexe, avec des aller-retour entre théories et observations, le tout saupoudré d'une dose de technologie et d'un grain de hasard (pour un court résumé, voir par exemple Chalmers 1976, « qu'est-ce que la science ? »). Même parmi les scientifiques, ce processus complexe n'est pas toujours correctement assimilé. Ainsi, le culte des héros, mâtiné d'un brin de révisionnisme, continue de faire florès malgré des études historiques poussées. Dans ce contexte, il peut être utile d'examiner de manière comparative la réaction à diverses observations, cruciales, l'interprétation de ces observations, et leur impact sur les théories développées alors. Quatre exemples sont présentés, tous liés à la question de la « construction des cieux » mais situés à diverses époques.

2. Antiquité : le modèle héliocentrique

Dans l'Antiquité grecque, les réflexions philosophiques firent apparaître des modèles d'univers « mécaniques », soit basés sur des principes « naturels » et non divins. Le premier exemple de cet effort sont les anneaux d'Anaximandre ($6^e$ siècle av. notre ère) : une Terre cylindrique est entourée de deux anneaux contenant du feu, et percé chacun d'un trou – l'un de taille inchangée, c'est le Soleil, l'autre de taille variable, c'est la Lune. Un siècle plus tard, Philolaos propose un modèle non géocentrique. Le centre du cosmos est un feu central autour duquel tourne le reste de l'Univers. Le Soleil y est un miroir en orbite réfléchissant le feu central ; une Terre plate orbite ce dernier en lui présentant toujours sa face arrière (le feu central est donc invisible depuis la face habitée de la Terre) ; une anti-Terre, orbitant en anti-phase avec notre maison cosmique, en contrebalance la masse. Le siècle suivant verra le triomphe du modèle géocentrique, soutenu par Platon, Eudoxe, et Aristote, où une Terre sphérique et immobile se place au centre du monde et est entourée de sphères célestes en rotation soutenant les astres connus (les 7 « planètes » et les étoiles).

Tous ces modèles étaient purement théoriques et se confrontaient peu aux mesures – seules des caractéristiques globales étaient considérées (ex. phases de la Lune, présence de rétrogadations planétaires). Bien au fait des modèles arithmétiques mésopotamiens, Hipparque connaît la grande précision de ceux-ci et il tente d'adapter les modèles grecs pour qu'ils « reproduisent les phénomènes », eux aussi. Ptolémée ira plus loin encore dans cette démarche, quitte à violer les préceptes d'Aristote pour ce faire (l'équant étant le meilleur exemple de cette démarche – les astronomes arabo-musulmans médiévaux arriveront finalement aux $13^e$ et $14^e$ siècle à éliminer ces ajouts non-aristotéliciens, mais au prix d'épicyles supplémentaires).

Toutefois, avant Hipparque, les observations avaient permis de définir au $3^e$ siècle un autre modèle. Le but initial de son concepteur, Aristarque, était de prendre la mesure des astres, en particulier les deux plus évidents : le Soleil et la Lune. Pour ce faire, il utilise leurs phénomènes les plus évidents : phases et éclipses. La première considération concerne une éclipse de Lune. Sachant que la totalité dure environ une heure et que la Lune se meut, par rapport aux étoiles, d'un angle équivalent à son diamètre en un heure, on trouve que la taille de l'ombre traversée par la Lune lors des éclipses a une taille de deux diamètres lunaires. Puisque cette ombre est portée par la Terre, on peut alors trouver le rapport des tailles des deux objets. Pour commencer, Aristarque suppose (une approximation qu'il lèvera plus tard) que le Soleil est très loin : l'ombre de la Terre étant alors très allongée, de forme quasi cylindrique. Sa taille valant donc un diamètre terrestre, Aristarque en déduit que la taille de la Lune est moitié de celle de la Terre. Connaissant la taille angulaire de la Lune[1], il peut alors déduire la distance Terre-Lune. Passant ensuite aux phases de la Lune, il remarque que la durée entre nouvelle Lune et premier quartier n'est pas égale à celle entre premier quartier et pleine Lune. Cela s'explique si le Soleil ne se trouve pas à distance quasi-infinie, comme considéré auparavant. En fait, au premier quartier, Terre-Lune-Soleil forment un triangle rectangle (avec l'angle de 90° en la Lune), ce qui permet de déduire le rapport des distances Terre-Soleil et Terre-Lune – un facteur entre 18 et 20, selon lui. Cela donne également le rapport entre les tailles des deux objets, puisque les deux possèdent la même taille angulaire, vu depuis la Terre (cela se remarque bien lors des éclipses solaires totales, où la Lune vient recouvrir parfaitement le Soleil). Cela pourrait suffire, mais Aristarque retourne aux éclipses lunaires, en supposant cette fois le Soleil à distance finie et l'ombre de la Terre conique, des hypothèses plus réalistes au vu des résultats précédents. Un simple calcul de proportions donne alors $1/s+1/l=1+o/l$, avec s le rayon solaire, l le rayon lunaire, et o la taille de l'ombre de la Terre à la distance de la Lune, le tout en rayons terrestres. Avec ses valeurs (taille angulaire observée, durée des éclipses, rapport des distances et des tailles Soleil/Lune), Aristarque trouve alors une Lune trois fois plus petite que la Terre et un Soleil presque sept fois plus grand que celle-ci.

Bien sûr, les données d'Aristarque sont imparfaites, et la moindre erreur a des conséquences fâcheuses sur la taille du Soleil (qui est en fait cent fois plus grand que la Terre et non sept). Néanmoins, même avec les mesures imparfaites de cette époque, impossible d'avoir une Terre dominante. La conséquence est, pour Aristarque, évidente : la Terre ne peut être le centre du Système, c'est l'objet majeur qui doit l'être, soit le Soleil. Il professe donc une théorie héliocentrique. Cette théorie ne sera pas vraiment bien accueillie : mis à part les accusations d'impiété, on opposera à Aristarque toute une série d'objections – que l'on retrouve dix-huit siècles plus tard (voir section suivante). Si certaines de ces objections sont tout à fait valides, il est piquant de constater que l'argument de la grande taille du Soleil, pourtant loin d'être anodin, disparaîtra quasi-totalement des débats ultérieurs... alors que le nom d'Aristarque sera bien mentionné. Le caractère décisif de l'observation, pour Aristarque, ne possédera pas le même attrait pour ses contemporains et successeurs.

3. Renaissance : le modèle hybride géohéliocentrique

La révolution scientifique des 16$^e$ et 17$^e$ siècles est souvent présentée comme le triomphe de la raison sur l'obscurantisme, les instruments optiques jouant ici un rôle-clé dans le renversement de l'ordre du monde. Si l'on examine les choses en détail, on se rend compte que la rationalité supposée est moins évidente dans les faits, et que la théorie la plus « rationnelle » n'est pas toujours celle que l'on pense.

Que démontrent en effet les fameuses premières observations au « télescope »[2] ? Trois choses distinctes, en fait. Tout d'abord, l'imperfection – pour utiliser un vocabulaire aristotélicien – des objets célestes : le Soleil présente des taches, la Lune des montagnes. Cela rend leur nature profonde apparemment plus proche de la matière terrestre que d'une matière éthérée, sans toutefois démontrer cette similitude formellement (il faudra attendre le 19$^e$ siècle et la spectroscopie pour pouvoir le faire).

---

1   Il prend un quinzième de signe, soit deux degrés, et non la taille réelle d'un demi-degré. Ce genre de détail fait penser à certains qu'il n'a pas mesuré lui-même tous les paramètres qu'il utilise et que le raisonnement tient plus de l'idée que de l'observation. Vu le peu d'écrits originaux d'Aristarque survivants, il est aujourd'hui difficile d'en juger.
2   La langue anglaise utilise telescope pour tous les instruments optiques astronomiques, mais la langue française fait la distinction entre les instruments avec lentilles (lunettes) et avec miroirs (télescopes). Au tournant du 17e siècle, c'est évidemment la lunette qui a conduit au débat décrit ici. Au passage, notons que Galilée n'est ni l'inventeur de cet instrument, ni le premier à s'en servir pour des observations astronomiques (ex. Nazé 2009, "Histoire du télescope" et références s'y trouvant).

Ensuite, la rotation de Vénus autour du Soleil : la planète présente des phases, et des variations de taille, très différentes de ce que l'on attend d'un modèle géocentrique. Chez Ptolémée, en effet, Vénus a au plus la forme d'un croissant fin et ne se présente jamais sous une forme gibbeuse ou pleine, puisque la planète ne passe jamais derrière le Soleil, vu de la Terre. (Figure : Phases de Vénus dans deux sysèmes du monde. http://www.astronomy.ohio-state.edu/~pogge/Ast161/Unit3/Images/venusphases.gif ) Enfin, la rotation de lunes autour de Jupiter. Ces deux dernières observations démontrent l'existence d'au moins deux centres de rotation en plus de la Terre, le Soleil et Jupiter, ce qui met à mal les bases de la physique aristotélicienne qui suppose un centre unique de rotation.

Ces observations permettent donc de rejeter le système ptolémaïque, mais elles n'apportent cependant pas de preuve formelle au système héliocentrique. Cela peut sembler étrange, puisqu'on suppose n'avoir le choix qu'entre ces deux théories. On oublie cependant que démontrer qu'une chose est fausse ne prouve qu'une autre est nécessairement vraie que si les deux sont mutuellement exclusives et forment l'ensemble des possibilités. Dans ce cas-ci, théories héliocentrique et géocentriques sont bien opposées, mais pas parfaitement exclusives. Il existe en effet au moins une autre théorie alternative : la théorie géohéliocentrique.

Cette théorie a été développée par Tycho Brahe, avant même les bouillonnements scientifiques du début du $17^e$ siècle (Blair A., 1990, Journal for the History of Ideas, 51, #3, 355-377). À son époque, les théories de Copernic étaient certes publiées, mais toujours pas démontrées, même en partie. Bien sûr, le système avait une certaine élégance, et permettait d'expliquer « naturellement » pourquoi les orbites planétaires étaient asservies au Soleil (ex. Vénus ne quittant jamais le giron solaire). Tycho le reconnaît sans détour. Par contre, ses objections au système sont assez nombreuses. Parmi celles-là, des oppositions classiques, datant de l'Antiquité : le fait de ne pas « sentir » la Terre bouger de quelque manière que ce soit – on ne perçoit pas de grands vents, on ne voit pas d'oiseaux devant lutter pour rattraper le sol qui fuit sous leurs pattes, on observe un objet lancé d'une haute tour tomber à ses pieds et non un peu plus loin ; ainsi que le problème d'avoir un objet énorme et bien solide à faire tourner, par une action qui doit être continue[3]. Une autre objection, éminemment pratique, existe aussi : le système de Copernic se voulait plus simple, ce qui était le cas pour certaines choses – le lien intime entre mouvement planétaire et Soleil – mais pas pour toutes : ainsi, pour reproduire les observations précisément, il fallait de nombreuses épicycles... et le centre du Système était le Soleil moyen, pas le Soleil réel.

Outre les raisons évoquées ci-dessus, Tycho Brahe possède aussi des raisons d'astronomie observationnelle (Blair A., 1990, Journal for the History of Ideas, 51, #3, 355-377). Ainsi, il commence par considérer Mars. Dans le système ptolémaïque, Mars se trouve toujours plus loin que le Soleil ; dans le système copernicien, Mars se trouve parfois plus près de la Terre que le Soleil, et ce lors des oppositions. Il suffit donc de mesurer la distance de la planète rouge pour choisir entre les deux modèles. Une telle distance peut en fait se mesurer directement, grâce au phénomène de parallaxe : un objet vu de deux endroits différents sera vu à des positions différentes par rapport à un arrière-plan – le changement angulaire de position est la parallaxe, et il est d'autant plus faible que la distance est grande. Cela s'applique à diverses situations : nos deux yeux pour la vision stéréoscopique, deux stations géodésiques pour la triangulation, deux positions sur Terre au même moment ou une position sur Terre en deux moments différents pour la parallaxe diurne, ou deux positions de la Terre sur son orbite autour du Soleil pour la parallaxe annuelle. En l'observant à l'opposition, Tycho estime donc de cette manière la distance de Mars, et il trouve une valeur de parallaxe inférieure à celle couramment acceptée à l'époque pour la parallaxe solaire : Mars se trouve donc apparemment plus loin que le Soleil, quand il devrait être proche, ce qui contredit le système copernicien.

---

3   Rappelons qu'à l'époque, tout mouvement devait être entretenu sous peine de s'éteindre rapidement. Cette conception du mouvement perdurera jusqu'aux travaux de Galilée et ses collègues. On montrera alors qu'un mouvement continue sur sa lancée, sans besoin d'action permanente pour l'entretenir. Soulignons aussi que les cieux étaient alors supposés constitués d'une matière différente de la matière terrestre solide, peu dense, ce qui facilitait la mise en mouvement, au contraire de la Terre "lourde".

En outre, ayant observé des comètes, Tycho ne leur trouve pas de mouvement rétrograde, contrairement aux planètes : cela suggère que ceux des planètes leur sont propres, et ne proviennent pas d'un effet de perspective dû au mouvement planétaire (en ce compris la Terre) autour du Soleil.

Enfin, Tycho dispose du meilleur observatoire occidental. Il peut donc observer précisément les positions des étoiles, en quête d'une quelconque parallaxe annuelle, mais il n'en détecte aucune. Copernic avait prévu cette objection, car l'absence de parallaxe annuelle était connue : il se disait qu'il suffirait que les étoiles se trouvent loin… Mais l'ampleur du problème lui échappait. La précision de Tycho (limite d'une minute d'arc) lui permet en effet de conclure que les étoiles doivent se trouver au moins à 700 fois la distance Soleil-Saturne (Siebert H., 2005, Journal for the History of Astronomy, 36, 251-271). Saturne étant la dernière planète du Système solaire connue à l'époque, il y aurait donc un énorme vide dans le cosmos… Pire encore : Tycho a également estimé la taille angulaire des étoiles – un quinzième du diamètre angulaire de la Lune (ou du Soleil) pour celles de première magnitude. En supposant qu'elles se trouvent juste derrière Saturne, comme dans le système ptolémaïque, leur taille vaut alors 80 % de celle du Soleil, ce qui semble raisonnable pour « d'autres soleils » brillant par leur propre lumière. Par contre, si elles sont 700 fois plus loin dans le système copernicien, elles doivent aussi être 700 fois plus grandes ! Les étoiles seraient donc des astres énormes, sans commune mesure avec le Soleil.

C'est pour tout cela que Tycho élabore dans les années 1580 son système hybride, compromis entre Charybde copernicienne et Scylla ptolémaïque (Figure : modèle tychonien http://en.wikipedia.org/wiki/Heliocentrism#mediaviewer/File:Tychonian_system.svg ). Dans ce système, les planètes tournent autour du Soleil, mais le Soleil lui-même tourne autour de la Terre. Mathématiquement et observationnellement, ce modèle est équivalent au système copernicien pour ce qui concerne les planètes. Par contre, la Terre reste immobile au centre de l'Univers dans le système tychonien, ce qui évite tous les écueils coperniciens. Il s'agit d'un compromis, qui garde les avantages en éliminant les inconvénients[4]. À noter que ce modèle présuppose que les planètes ne se baladent pas sur des orbes physiques, solides et réelles, car les orbites s'y entremêlent : cela ne pose aucun problème car Tycho lui même avait démontré leur inexistence, en observant que la comète de 1577 se baladait entre les planètes et aurait donc transpercé les sphères célestes, si elles existaient.

Cependant, dans les années qui suivent, certains problèmes seront réglés. Ainsi, l'étude du mouvements des corps fait naître la notion de « mouvement commun » et d'inertie : un objet tombant du mât d'un navire en translation possède, outre le mouvement de chute, le mouvement du bateau, ce qui explique qu'il atterrit au pied du mât ; d'autre part, un objet en mouvement le conserve tant qu'il n'y a pas de frottement. De plus, Kepler résout l'écueil des épicycles coperniciennes avec l'utilisation d'orbites elliptiques et non plus circulaires. Reste néanmoins le cœur du système copernicien : les mouvements de la Terre – mouvement diurne, soit de rotation sur elle-même, et mouvement annuel, soit de rotation autour du Soleil[5]. C'est sur la (non-) détection de ces mouvements que se cristallise le débat.

Reprenons le mouvement annuel. Bien sûr, après Tycho, les observations ont changé, grâce à la lunette. L'image des étoiles dans l'instrument optique est bien plus petite que celle estimée à l'œil nu. Cela pourrait résoudre le problème de taille immense… ne serait le fait que la lunette permet des mesures plus précises, ce qui diminue encore les limites sur les valeurs de parallaxe stellaire. Les étoiles se voient donc reléguées plus loin encore, ce qui ne compense pas la diminution de leur taille apparente : les astres restent énormes comparés au Soleil. La seule réplique des coperniciens à ce sujet repose sur la foi : Dieu peut faire des astres aussi grands qu'il le désire (Graney C.M. 2013, Journal for the History of Astronomy, 44, 165-172). Cette objection bien peu rationnelle ne sera pas suivie par tous. Ainsi, Simon Marius se targue de suivre la logique jusqu'au bout dans son *Mundus Jovialis* (1614), privilégiant la système tychonien car les étoiles, puisqu'elles sont résolues dans la lunette, ne peuvent être lointaines (Graney C.M. 2010, *Physics in Perspective,* 12, #1, 4-24, et arXiv:0903.3429).

---

4   Il faut cependant insister sur l'équivalence du système tychonien au copernicien pour les planètes : les objections de Tycho à propos de Mars et des comètes s'appliquent donc également à son modèle… En fait, pour Mars, son erreur repose sur l'acceptation de la parallaxe solaire, dont la valeur était alors grossement surestimée.

5   Chez les premiers coperniciens, il y avait un troisième mouvement, qualifié "de libration" ou "de précession": on ne savait pas à l'époque qu'un objet en rotation est stable, son axe restant fixe dans l'espace. Du coup, pour que la Terre maintienne son axe de rotation fixe dans l'espace lors de son mouvement autour du Soleil, il fallait imposer un mouvement supplémentaire.

Galilée lui-même est bien conscient du problème que constitue la parallaxe. Dans son *Dialogo sopra i due massimi sistemi del mondo* (1632)[6], il propose une nouvelle observation : regarder un couple d'étoiles de magnitude différentes car selon toute logique, l'une doit alors se trouver près de l'observateur, l'autre loin. Elles possèdent donc des parallaxes différentes, soit une parallaxe différentielle facilement détectable selon lui – cette « simple » observation devrait clôturer le débat. Toutefois, ce qu'il oublie de mentionner, c'est qu'il a fait cette expérience, avec des résultats négatifs (Siebert H., 2005, Journal for the History of Astronomy, 36, 251-271 ; Graney C.M., arXiv:physics/0606255) ! Dès 1617, il a observé Mizar, dont les deux composantes présentent des luminosités différentes. Des tailles observées, il déduit que Mizar A se trouve à 300 fois la distance Terre-Soleil, et Mizar B à 450 fois cette valeur, et que la position relative des deux objets doit donc varier de plusieurs minutes d'arc, un changement facilement détectable. Hélas, il n'en est rien, et Galilée a beau reproduire l'observation avec divers autres groupes stellaires (dont le Trapèze d'Orion), il ne trouve toujours aucune validation du système copernicien – au contraire, il s'agit d'une réfutation ! Le fait qu'il ait soigneusement caché cette multiple non-détection montre qu'il était conscient de l'importance de l'objection qu'elle constituerait, et souligne aussi son attitude parfois hypocrite (assurer dans son *Dialogo* qu'il suffit de faire l'expérience alors qu'il l'a faite avec un résultat négatif !) et peu scientifique...

Après Tycho et Marius, de nombreux jésuites reprennent le flambeau de la défense du système hybride (ex. Cristoforo Borri, Carolino L.M. 2008, Journal for the History of Astronomy, 39, 313-344), l'exportant même jusqu'en Chine. Le plus connu d'entre eux est Riccioli. Dès le frontispice de son *Almagestum Novum* (1651) (Figure : Frontispice de l'Almagestum Novum de Riccioli http://en.wikipedia.org/wiki/Giovanni_Battista_Riccioli#mediaviewer/File:AlmagestumNovumFrontispiece.jpg ), le ton est donné : le système ptolémaïque gît au sol, éliminé par les avancées scientifiques, en balance restent le système héliocentrique et le système géohéliocentrique[7], qui est favorisé. Dans son livre, Riccioli recense 126 arguments (et contre-arguments) à portée cosmogonique, 49 pour le système copernicien et 77 contre ce dernier (Graney C.M., 2012, Journal for the History of Astronomy, 43, 215-226 et arXiv:1103.2057). Cette profusion s'explique par un souci de complétude, pas par la valeur des arguments eux-mêmes : ainsi, il rejette la majorité de ces arguments car ils sont peu persuasifs voire ineptes. D'autre part, il n'insiste pas sur le côté religieux, qui n'occupe que deux argument « contre » (et deux arguments rejetés par lui!), et le contre-argument sur la taille des étoiles (comme déjà évoqué plus haut), là aussi rejeté par Riccioli. Au final, il ne lui reste que quelques arguments sans réponse valide, tous du côté anti-copernicien.

Si l'on exclut les arguments de simplicité ou de manque de système cohérent, il reste les arguments décisifs, basés sur l'observation. Contre le mouvement annuel de la Terre, Riccioli reprend en détail le problème d'absence de parallaxe détectable et sa conséquence sur la taille des étoiles. Contre le mouvement diurne, il insiste sur l'absence de preuve de mouvement de rotation (dont Tycho avait aussi eu l'intuition). En effet, Galilée et ses acolytes pensaient avoir résolu tous les problèmes avec son « mouvement commun », mais ce dernier ne s'applique que pour des mouvements de translation pure, or la Terre est en rotation... Riccioli montre que ce mouvement de rotation a des conséquences observables, mais non détectées. Ainsi, un objet lancé d'une haute tour posée sur une Terre en rotation ne devrait pas atterrir exactement à la verticale du point de chute. D'autre part, si un canon tire un boulet vers le nord ou le sud, la différence avec la latitude de la vitesse associée à la rotation provoque une déflexion, que Riccioli pensait détectable mais savait non rapportée par les artificiers de l'époque. Enfin, si un boulet est tiré vers l'est ou l'ouest, des déflexions doivent également apparaître, mais aucun n'est là aussi rapporté. En tout cela, Riccioli est un précuseur : Newton ré-imaginera l'expérience de la chute quelques décennies plus tard, Laplace en 1778 et Coriolis en 1835 détailleront les déflexions dans un système en rotation (qui s'applique au cas nord/sud), tandis que Eötvös mettra en évidence l'effet est/ouest au début du $20^e$ siècle.

---

6   À noter que ce "dialogue" évite soigneusement le troisième larron, le système tychonien. Forcément, diront certains, vu que Galilée ne possède pas d'objection réelle contre lui...

7   À noter que le système représenté n'est pas parfaitement tychonien. En effet, si Mercure, Vénus, et Mars tournent autour du Soleil, Jupiter et Saturne tournent autour de la Terre. Ces deux planètes possédant des satellites, au contraire du trio précité, Riccioli en faisait des mondes à part entière. Les satellites de Mars ne seront découverts qu'en 1877.

Ces deux grandes objections observationnelles (absence de déflexions, absence de parallaxe) auraient dû faire rejeter le système héliocentrique, falsifié, au profit du système hybride. C'est ce que fait Riccioli, mais pas ses contemporains. Par la suite, les commentaires sur son travail se feront au mieux moqueurs, au pire dénigrants (Graney C.M., 2012, Journal for the History of Astronomy, 43, 215-226 et arXiv:1103.2057), et ce alors qu'il n'avait fait que suivre la méthode scientifique et que l'on ne pouvait alors accepter le système copernicien que par un acte de foi.

Les véritables preuves du mouvement de la Terre attendront en effet quelques années encore. Le mouvement annuel sera mesuré via le phénomène d'aberration par Bradley en 1727, et la parallaxe stellaire sera détectée dans les années 1830 (Bessel avec 61 Cyg, Henderson avec Alpha Cen, Struve avec Véga). Le mouvement diurne sera observé via la forme aplatie de la Terre mesurée lors des expéditions françaises au Pérou et en Laponie dans les années 1730, la déflexion Coriolis d'un objet tombant dans un puits de mine profond au début au $19^e$ siècle, et le mouvement du pendule de Foucault (1851). Si Horrocks puis Halley montrèrent que la taille des étoiles était probablement plus faible qu'on ne le pensait, car les étoiles disparaissaient de manière quasi-instantanée lors des occultations par la Lune, il fallut attendre le développement des théories optiques pour véritablement comprendre que la « tache » stellaire observée par l'œil ou le télescope ne correspond pas à la taille angulaire réelle de l'astre mais dépend de l'optique (phénomène de la tache d'Airy, 1835) : la première véritable mesure d'un diamètre stellaire sera fait par interférométrie au tournant des années 1920.

4. Lumières : notre environnement stellaire

Avec l'acceptation du modèle héliocentrique, les astronomes voient le Soleil comme une étoile parmi d'autres : la « sphère des fixes » disparaît, les étoiles se dispersant dans l'espace. Reste à savoir comment elles s'organisent. Du côté théorique, Newton considère l'univers infini et les étoiles qui s'y trouvent distribuées uniformément. En effet, un univers fini ou une légère anicroche à l'uniformité provoquerait un effondrement gravifique, une catastrophe touchant l'ensemble de l'univers. C'est la vision partagée par la plupart de ses contemporains, et de ses successeurs. Toutefois, ce modèle pose plusieurs problèmes (Norton J.D., 1999, in The expanding worlds of general relativity, p271-322, Hoskin M., 2008, Journal for the History of Astronomy, 39, 251-264). Tout d'abord, l'équilibre parfait dans un univers infini, isotrope, et uniforme suppose que les forces se compensent exactement dans toutes les directions (soit attraction infinie perçue d'un côté moins attraction infinie reçue de l'autre valant zéro), alors que la réponse mathématique est en réalité indéterminée. Newton lui-même n'accorde que peu de crédit à cette objection, mais il échoue toutefois à la résoudre, faisant appel à l'intervention stabilisatrice de Dieu en cas de problème local. D'autre part, un univers infini suppose un nombre infini d'étoiles... et donc un ciel brillant ! Ce paradoxe est connu sous le nom de paradoxe d'Olbers. Tout comme le problème d'effondrement (quoique ce dernier soit moins « médiatisé »), il sera longuement discuté, tout au long des $18^e$ et $19^e$ siècles, ainsi qu'au début du $20^e$ siècle (moment où ils seront enfin résolus).

Enfin, il y a aussi l'aspect du ciel, qui est loin d'être uniforme : une large structure, la Voie Lactée, traverse en effet la voûte céleste de part en part. Cet objet offre une autre possibilité de structuration pour l'Univers. En effet, les observations à la lunette ou au télescope montrent que la Voie Lactée est composée de nombreuses étoiles, ainsi que les atomistes le supposaient. Au $18^e$ siècle, on explique donc simplement l'apparence du ciel en supposant que le système solaire se trouve au sein d'une « couche » d'étoiles (une coquille sphérique pour Thomas Wright, un disque plat pour Kant et Lambert), qui n'est d'ailleurs peut-être pas unique – c'est l'hypothèse bien connue des « univers-îles » (du nom donné par von Humbolt). Toutefois, ces propositions restent spéculatives et, au mieux, qualitatives.

Cela va changer avec les travaux de William Herschel sur la « construction des cieux » (Hoskin M., 2012, The Construction of the heavens, Chaberlot F., 2003, La Voie Lactée). Pour fixer quantitativement la forme de notre Voie Lactée, Herschel va procéder au comptage des étoiles, inventant au passage la statistique stellaire. Pour transformer ces nombres en contraintes structurelles, il doit évidemment faire quelques hypothèses : (1) la distribution des étoiles est uniforme jusqu'aux limites du système, (2) les étoiles sont toutes semblables entre elles, (3) l'univers est fini et accessible à l'observation. Les deux premières hypothèses permettent d'affirmer que si l'on voit plus d'étoiles dans un champ, c'est simplement que l'on voit plus loin, donc que la limite de l'univers est plus éloignée, dans cette direction : le nombre d'astres, N, est proportionnel au cube de la distance, $r^3$. Observant 3400 champs, une tâche fastidieuse mais nécessaire, Herschel détermine la distance de la limite de l'univers dans ces directions. En 1782-1785, il parvient ainsi à la conclusion que la limite la plus éloignée se trouve à 498 fois la distance Sirius-Soleil[8] et il utilise ses résultats pour dresser une carte des frontières de notre système sidéral (Figure : Structure de la Voie Lactée, selon une coupe passant par les pôles galactiques, selon Herschel http://commons.wikimedia.org/wiki/File:Herschel-Galaxy.png ).

Dans les années qui suivent, Herschel découvre cependant plusieurs problèmes dans son raisonnement. Son télescope de 40 pieds montre en fait plus d'étoiles que celui de 20 pieds, utilisé pour le comptage : cela sous-entend que la limite de l'Univers ne lui était alors pas accessible auparavant, et ne l'est d'ailleurs peut-être pas encore. En outre, ses recherches de « nébuleuses » lui ont fait découvrir de nombreux agrégats d'étoiles : la distribution stellaire est donc loin d'être uniforme. Il résume ses vues en 1817-1818, assurant la Voie Lactée insondable. Toutefois, il refuse de rejeter l'hypothèse de similitude des étoiles et la relation magnitude-distance, même s'il a découvert que (1) de nombreuses étoiles doubles (dont il a fini par comprendre la véritable nature – une paire physiquement liée et non une simple coïncidence) sont asymétriques, remettant en cause la luminosité universelle des étoiles et (2) la distribution des étoiles en fonction de leur magnitude, dans le catalogue de Bode, ne correspondait pas à ses attentes basées sur une proportion magnitude-distance.

La révision viendra de ses successeurs, qui relaxent les hypothèses de distribution uniforme des étoiles et de similitude stellaire et utilisent la paramétrisation correcte des magnitudes (loi logarithmique et non linéaire). Toutefois, l'insondabilité finalement conclue par Herschel est oubliée et sa carte galactique, pourtant reniée, est reproduite abondamment. Il y a là un paradoxe intéressant : l'utilisation enthousiaste d'une technique dont l'inapplicabilité a pourtant été démontrée. L'aboutissement de ces (vains?) travaux de continuation sera atteint chez Hugo von Seeliger et Jacobus Kapteyn, qui obtiennent une Voie Lactée en sphéroïde aplati, de taille moyenne (10 à 20kpc) et avec un Soleil quasi-central (occupant donc une place privilégiée). La solution de Kapteyn, en particulier, est très populaire, au point de prendre le sobriquet d' « univers de Kapteyn ».

---

8 Par ailleurs, selon Herschel, le rapport des magnitudes de deux astres est proportionnel au rapport de leur distance : rapporter une distance de 498 fois Sirius-Soleil implique donc qu'il observe des étoiles de la $498^e$ magnitude, un nombre clairement invraisemblable. Herschel ne se rendit cependant pas compte du problème (ou le passa sous silence). Notez que l'œil nu peut voir jusqu'à la sixième magnitude alors que les meilleurs télescopes actuels descendent jusqu'à la $30^e$ magnitude.

Au 19ᵉ siècle coexistent donc deux modèles cosmiques – univers infini et uniforme, Voie Lactée de taille finie et centrée sur le Soleil (objet unique ou entouré d'autres univers-îles, selon les auteurs et les époques) – sans véritable concurrence entre les deux, malgré leurs oppositions. Les problèmes de l'un (effondrement gravitationnel catastrophique, paradoxe d'Olbers) et de l'autre (accessibilité de la frontière, héliocentrisme galactique peu copernicien) sont connus mais restent non résolus, et finalement peu débattus. La crise sous-jacente ne pouvait qu'être reportée, nécessitant d'abord les choses d'un angle neuf, comme le montre la section suivante.

5. Vingtième siècle : le grand débat de 1920

L'Univers ne se compose pas seulement de planètes et d'étoiles, mais ces objets-là ont largement dominé l'astronomie jusqu'à la fin du 19ᵉ siècle. À l'œil nu, on ne peut repérer que quelques autres structures : la Voie Lactée, bien sûr, mais il y a aussi les Nuages de Magellan (visibles depuis l'hémisphère sud) et les « nébuleuses » d'Orion et d'Andromède, parmi les plus célèbres. Jusqu'au 16ᵉ siècle, on ne connaît qu'une poignée de nébuleuses, mais le nombre des objets « flous » croît fortement avec l'utilisation des télescopes et des lunettes et, à la fin du 18ᵉ siècle, Charles Messier en catalogue une centaine, tandis que le travail systématique de Caroline et William Herschel dévoile 2500 nébuleuses.

La question de leur nature se pose immédiatement. Certaines nébuleuses se révèlent être, dans les télescopes puissants d'Herschel, des amas d'étoiles, un peu comme la Voie Lactée avait été résolue en étoiles par la lunette de Galilée. Herschel pense donc au départ que toutes les nébuleuses sont des groupes stellaires, et que celles qui restent floues attendent simplement l'avènement d'un plus gros instrument pour révéler leur vraie nature. Toutefois, dans les années 1780-1790, il découvre de véritables nuages célestes, irréductibles en étoiles – des objets qu'il appelle « nébuleuses planétaires »[9] - et il doute alors d'une nature unique pour l'ensemble des nébuleuses.

Cependant, une théorie unique reste privilégiée par les astronomes, tout au long du 19ᵉ siècle et ce malgré les observations. À cette époque, les faits s'accumulent en effet, mais sont diversement appréciés. Lord Rosse découvre avec son télescope « Léviathan » la forme spirale de M51 (51ᵉ objet du catalogue de Messier), et James Keeler démontre un demi-siècle plus tard qu'une large fraction des nébuleuses possède cette forme. Certains y voient une preuve qu'il s'agit d'étoiles, d'autres qu'il s'agit de gaz au premier stade de la formation stellaire. William Huggins et son collègue William Miller (en 1864) puis Julius Scheiner (en 1899) analysent le spectre de plusieurs nébuleuses : ces pionniers trouvent que certaines présentent un spectre solaire (lignes noires sur fond brillant) mais d'autres un spectre de gaz chaud (lignes brillantes sur fond noir) – la dichotomie imaginée par Herschel se confirme donc, mais est peu appréciée au moment même. La découverte de « novae » brillantes en 1885 dans M31 (S And) et en 1895 dans NGC5253 (Z Cen) semble aussi rejeter l'hypothèse d'univers-îles indépendants, puisque ces objets auraient une brillance extraordinaire s'ils étaient loin. Enfin, les positionnements présentent des différences marquées : les amas stellaires dits « ouverts » et les nébuleuses planétaires se trouvent dans le plan de la Voie Lactée, les amas stellaires dits « globulaires » s'affichent en dehors de ce plan, mais pas trop loin de lui, tandis que les « nébuleuses spirales » s'en éloignent fortement. S'il semble évident de considérer les deux premiers comme partie intégrante de la Voie Lactée, le lien est plus discuté pour les autres (l'évitement implique-t-il toujours une association ?), et les conséquences sur leur nature sont moins évidentes encore.

La nature exacte des nébuleuses floues, irrégulières ou spirales, restait donc à éclaircir, tout comme leur distance (objets proches et gazeux, univers-îles lointains et stellaires, Fernie J.D., 1970, Publications of the Astronomical Society of the Pacific, 82, 1189). Tout cela faisait l'objet de nombreuses discussions à l'époque, avec un balancier oscillant régulièrement entre les deux interprétations extrêmes. La résolution du conflit va venir d'une direction inattendue car c'est en étudiant les amas globulaires qu'un astronome fraîchement diplômé, Harlow Shapley, va créer la polémique.

---
9   Le nom provient du fait qu'Herschel y voyait un nuage en train de se condenser en planètes. On sait aujourd'hui qu'il s'agit au contraire du dernier stade évolutif des étoiles de type solaire – une trace de mort plus que de naissance, donc – mais le nom est resté.

Entre 1915 et 1921, il les observe systématiquement, pour en déterminer la distance. Il utilise trois méthodes. La première s'applique aux amas proches, et repose sur les variations des Céphéides et des RR Lyrae. Henrietta Leavitt avait découvert quelques années plus tôt que la période de variation est corrélée à la luminosité absolue pour les Céphéides : une fois la période mesurée, il suffit dès lors de comparer luminosités apparente et absolue pour déterminer la distance. Pour ce faire, il faut évidemment disposer d'une relation période-luminosité calibrée, un travail que Shapley a justement effectué en utilisant des Céphéides galactiques. Pour des amas plus lointains, il ne peut distinguer les Céphéides, alors il se concentre sur les 30 étoiles les plus brillantes de l'amas (en retirant ensuite les 5 extrêmes, pour éviter toute contamination d'avant-plan) : en comparant leur luminosité apparente à la luminosité connue d'objets brillants proches, il trouve leur distance. Enfin, pour les amas les plus éloignés, il ne peut résoudre aucune étoile et mesure leur diamètre pour le comparer à celui d'amas proches, ce qui suppose une uniformité dans les propriétés de ces amas. Tout cela lui permet de former une carte 3D des amas, qui s'avère fortement asymétrique par rapport au Soleil. Supposant que ces amas sont partie intégrante de la Voie Lactée (une hypothèse non acceptée par tous !), il en déduit non seulement une taille énorme (300 000 années-lumières, soit dix fois plus qu'accepté à l'époque) pour celle-ci mais aussi une position fortement excentrée pour le Soleil.

Il s'agit d'une seconde révolution copernicienne, en quelque sorte, et après la première on pourrait croire à un accueil favorable... mais ce n'est pas le cas. Après des échanges houleux dans les revues scientifiques, un « débat » sur la taille de l'univers (« *The scale of the Universe* ») est finalement organisé le 26 avril 1920 par George Hale. Les guillemets sur débat sont de mise, car il n'y eut pas de véritable débat : chacun expose ses conclusions, sans échange d'aucune sorte[10]. Toutefois, un tel match reste inédit en histoire de l'astronomie et il arrive en plus à point nommé, ce qui explique son aura quasi-légendaire. Les « débatteurs » en présence sont d'un côté le jeune Harlow Shapley, alors inexpérimenté côté discours (il lira péniblement les 19 pages qu'il a préparées) et en lice pour obtenir un poste important, et de l'autre Heber Curtis, astronome senior et excellent orateur. Leur discours original (Hoskin M., 1976, Journal for the History of Astronomy, 7, 169) et la version imprimée des contributions (Shapley H., 1921, Bull. Nat. Res. Coun., 2, 171 et Curtis H., 1921, Bull. Nat. Res. Coun., 2, 194) ont tous deux survécu. Les discours étaient très courts et, dans le cas de Shapley, de niveau général, tandis que les versions imprimées, échangées entre les deux avant impression, rentrent dans plus de détails techniques. On peut résumer leurs modèles et leurs arguments comme suit[11] :

Modèles
- Shapley : la Voie Lactée, agrégat de groupes stellaires, a une taille de 300 000 années-lumière et le Soleil est positionné au moins à 50 000 années-lumière de son centre ; les nébuleuses spirales sont des objets proches et petits, qui lui sont subordonnés et appartiennent donc à un halo étendu de notre Galaxie.
- Curtis : la Voie Lactée a une taille de pas plus de 30 000 années-lumière, et le Soleil est

---

10 À noter que Shapley parla premier, Curtis second, mais que cela ne marqua pas la fin des choses. Des questions ou remarques du public étaient autorisées, et Henri Russell, le promoteur de thèse de Shapley et un adversaire de la théorie d'univers-îles, répondit ainsi aux objections de Curtis à la fin du "débat" (Shapley s'était évidemment assuré de sa présence). Hélas, il ne reste aucune trace du contenu exact de son discours.

11 Outre les références déjà citées : Shapley H., 1968, Through rugged ways to the stars ; Chaberlot F., 2003, La Voie Lactée ; Trimble V., 1995, Publications of the Astronomical Society of the Pacific, 107, 1133 ; Struve O., 1960, Sky & Telescope, numéro de mai, 19, 398 ; Hetherington N.S., 1970, Astronomical Society of the Pacific Leaflets, 10, #490, 313 ; Smith R.W., 1982, The expanding Universe ; Longair M., The cosmic century, 2006 ; Smith R.W. 2006 Journal for the History of Astronomy, 37, 307 et 2009, Journal for the History of Astronomy, 39, 91 (voir aussi son livre « The expanding Universe », 1982).

positionné près de son centre ; les nébuleuses spirales lui sont similaires et sont donc d'autres galaxies.

Arguments

- *Nature et luminosité des étoiles dans les amas globulaires.* Tout d'abord, Shapley a détecté grâce à leur couleur des étoiles bleues, chaudes, dans les amas globulaires, et il a confirmé leur nature spectroscopiquement. Il les estime donc être similaires à celles se trouvant dans la Voie Lactée : comme ces dernières sont des géantes brillantes, il détermine de grandes distances pour les amas globulaires. Curtis répond ici en mettant en avant l'incertitude quant à la similarité de ces objets et son désaccord quant aux magnitudes considérées comme typiques. Ensuite, Shapley assure que les étoiles les plus brillantes, jaunes et rouges, qu'il utilise pour déterminer les distances, sont aussi des géantes : en effet, d'un groupe lointain, on doit observer en premier les membres les plus brillants et d'ailleurs les amas proches sont clairement dominés par les géantes. Il compare donc la luminosité apparente des étoiles brillantes des amas globulaires à la luminosité, connue, des géantes dans les amas ouverts proches et obtient là aussi de grandes distances. De plus, le spectre des objets qu'il a pu étudier montre distinctement la signature d'étoiles géantes, ce qui le conforte dans ses conclusions. Au contraire, Curtis assure que les étoiles dans notre voisinage et dans les amas sont rarement des géantes mais généralement des naines, et donc que les étoiles des amas globulaires doivent être aussi, en majorité, des naines – par un argument copernicien qui veut que l'Univers soit partout similaire[12] : comme les naines sont bien moins brillantes intrinsèquement que les géantes, les amas globulaires sont plus proches que ce que Shapley avance. En fait, c'est Shapley qui a raison ici, même si les étoiles des amas globulaires se révéleront quelque peu différentes de celles des amas ouverts et si l'extinction joue un rôle dans la calibration des distances (voir plus bas). Il est à noter que Shapley avance un autre argument quant aux grandes distances des amas globulaires : leur vitesse importante, mesurée sur les spectres, devrait se traduire par un changement de position détectable s'ils étaient proches et comme on n'en détecte aucun, ils sont donc assez éloignés. Cela ne sera pas réfuté par Curtis.
- *Céphéides.* Shapley calibre sa relation en utilisant onze céphéides galactiques : ce n'est pas tant les valeurs individuelles qui l'intéressent mais la moyenne, qui suffit si la forme de la relation est calibrée par ailleurs. Curtis assure qu'une calibration avec onze objets est, au mieux, incertaine ; en outre, il ne voit nulle trace de relation période-luminosité dans les objets galactiques, et il en déduit que les Céphéides de la Voie Lactée et celles des amas diffèrent peut-être. Quoique Curtis ait raison sur les propriétés différentes des Céphéides dans les amas (elles sont en fait moins brillantes, cf. Population II ci-dessous), Shapley est correct quant à l'existence universelle de la relation période-luminosité (et sa méthode de calibration est valide et aurait fonctionné... si les luminosités galactiques avaient été corrigées de l'extinction).
- *Taille de la Voie Lactée.* Des grandes distances trouvées pour les amas globulaires, Shapley déduit sa grande taille pour la Voie Lactée. De la distribution asymétrique de ces amas (bien connue au niveau de la répartition sur la voûte céleste, mais encore plus évidente avec son modèle 3D combinant coordonnées célestes et distances), il déduit la position non centrale du Soleil. De son côté, Curtis se repose sur les travaux de statistique stellaire, qui trouvent une Voie Lactée petite et centrée sur le Soleil. C'est ici Shapley qui a raison, l'extinction faussant les comptages stellaires.
- *Nova et distance des spirales.* En 1917, Ritchey, Curtis, et Shapley ont détecté quelques novae dans des nébuleuses spirales. Supposant qu'il s'agit d'objets similaires aux quelques novae galactiques de distance connue, Curtis en déduit une distance assez élevée pour les spirales (500 000 années-lumière pour M31) ; à cette distance, la taille angulaire de ces objets se transforme en une taille réelle assez similaire à celle qu'il attribue à la Voie Lactée. Shapley, vu la taille de sa Voie Lactée et la faible distance des spirales, arrive évidemment au résultat inverse : les spirales sont petites, d'une taille pas du tout comparable à la Voie Lactée. Par contre, il ne parle pas des novae lors du débat. C'est Curtis qui a raison à ce niveau, même si la calibration galactique de sa distance est incorrecte à cause de l'extinction et si les deux novae de 1885 et 1895 posent alors un gros problème (Hoskin M., 1976, Journal for the History of Astronomy, 7, 47-53) – leur luminosité intrinsèque serait énormément plus élevée qu'observé dans la Voie Lactée notamment pour les novae classiques (il s'agissait en fait de « supernovae », ce qui sera reconnu un peu plus tard).

---

12  Ce qui est, en soi, assez piquant, vu que Curtis place le Soleil au centre de la Galaxie, une place bien peu copernicienne!

- *Mouvements des spirales.* Curtis précise que la grande vitesse radiale observée pour les nébuleuses spirales démontre leur nature extragalactique, vu que de telles vitesses ne sont pas observées pour les objets de la Voie Lactée. Shapley fait, lui, appel aux résultats de son collègue et ami Adriaan Van Maanen. Ce dernier avait rapporté le changement de position, faible mais non nul, dû à la rotation et à l'expansion de certaines nébuleuses spirales : pour que ce déplacement ait lieu à une vitesse inférieure à celle de la lumière, il faut que les objets soient proches. C'est Curtis qui a raison sur ce point, même s'il ne réalise pas la portée qu'auront les vitesses des spirales (voir ci-dessous). Les résultats de Van Maanen sont en fait faux[13] mais Shapley ne pouvait pas le savoir. Il est ici amusant de constater une inversion des opinions. Shapley avait des doutes au départ, vers 1917-1920, sur les résultats de Van Maanen (Berendzen R. & Hart R., 1973, Journal for the History of Astronomy, 4, 46-56). Par contre, Curtis avait rapporté l'existence de mouvements propres (pas de rotation et d'expansion, comme Van Maanen, mais un mouvement transversal de l'ensemble – la difficulté de détection est similaire au vu des chiffres en cause) pour les nébuleuses spirales quelques années plus tôt (Curtis H. 1915, Publications of the Astronomical Society of the Pacific, 27, 214) : cela impliquait des distances de quelques dizaines de milliers d'années-lumière pour ces objets, bien moins que ce qu'il trouve avec les novae, un détail qu'il passe sous silence en 1920 !

- *Apparence des spirales.* Curtis avance que le spectre des nébuleuses spirales est similaire à celui d'étoiles, et qu'ils doivent donc être des groupes d'étoiles, comme la Voie Lactée. Shapley utilise les résultats de collègues qui montrent que la distribution de couleur et luminosité, ainsi que la brillance de surface, des spirales diffèrent de ceux de la Voie Lactée. Il doit donc s'agir d'autres objets, peut-être nébuleux de nature[14]. Précisons cependant que Shapley n'élimine pas la possibilité d'autres galaxies, mais qu'il les situe bien plus loin. C'est cependant Curtis qui a raison sur ce point, les problèmes d'apparence que pointe Shapley proviennent de la non-correction de l'extinction dans la Voie Lactée.

- *Zone d'évitement.* Curtis a observé des nébuleuses spirales vu par la tranche, et y a découvert une zone sombre, qu'il interprète comme des nuages obscurcissant la vue. Si la Voie Lactée possède un tel « anneau » de nuages, alors, cela explique qu'on ne puisse voir de nébuleuses spirales dans le plan galactique, leur lumière étant totalement absorbée. Pour expliquer l'absence de spirales dans ce plan, Shapley avance plutôt une explication dynamique : il avait supposé que les amas globulaires se « rompaient » quand ils traversaient le plan galactique (ce en quoi il a raison, même s'il fallut bien longtemps pour prouver l'existence du phénomène), il suppose alors l'existence d'une force répulsive pour les nébuleuses spirales, qui permet par ailleurs d'expliquer leur grande vitesse. C'est Curtis qui a raison ici, sauf qu'il ne tire pas les conséquences de la présence d'extinction sur la fiabilité des statistiques stellaires, dont il reprend les résultats quant à la taille de la Voie Lactée.

---

13 Hubble puis van Maanen publieront en 1935 une réfutation de ses résultats antérieurs. Par contre, au début des années 1920, ils étaient souvent considérés corrects et abondamment utilisés dans le débat concernant l'existence d'univers-îles – et ce alors que les preuves contre eux s'accumulaient (Berendzen R. & Hart R., 1973, Journal for the History of Astronomy, 4, 73-98). Leur prestige fut tel que Hubble retarda d'un an sa publication de la découverte de Hubble de Céphéides extragalactiques (voir ci-dessous). C'est pourtant elle qui amorcera la critique ouverte des travaux de Van Maanen, dont l'influence sera minime après 1925.

14 Reprécisons ici qu'on a souvent eu tendance à la fin du 19e siècle et au début du 20e siècle à associer les nébuleuses aux premiers stades de formation stellaire, un peu à la manière d'Herchel (note 9) – il devait donc s'agir d'objets proches... Jeans étendra cette idée après 1917 à des spirales qu'il suppose formant un groupe d'étoiles (et non plus une seule comme proposé auparavant).

Les arguments décisifs de l'un comme de l'autre reposent sur des observations précises, même si pas toujours fiables. Seul problème : la totale non-communication. En effet, Shapley est spécialiste des amas globulaires tandis que Curtis est spécialiste des nébuleuses spirales : chacun parle de ce qui le concerne (étoiles pour l'un, spirales pour l'autre) et, l'astronomie s'étant spécialisée à l'époque, chacun ignore (presque) tout du domaine de l'autre et en parle donc de manière moins convaincante. Shapley doit ainsi faire appel, sur le sujet qu'il ne maîtrise pas, aux résultats de ses collègues (qui se révéleront parfois faux) ; il faut aussi reconnaître qu'il est un peu hâtif sur certains points (comme la nature universelle des Céphéides). Curtis, lui, utilise l'argument de prouver la similarité et de confirmer les calibrations de manière sélective : il exige plus d'observations de la part de Shapley alors qu'il se contente lui-même de quelques objets pour calibrer ses propres résultats. L'issue en est très logiquement un match nul : Shapley avait globalement raison en ce qui concerne les amas, Curtis pour les spirales. Par contre, aucun des deux modèles d'univers n'a survécu longtemps à ce débat. Les décennies suivantes vont en effet profondément modifier notre vision de la structure de l'Univers.

Ainsi, l'extinction, négligée par tous jusqu'alors, va se révéler être bien réelle. Cette extinction provoque une absorption de la lumière et un rougissement de celle-ci. Curieusement, cet effet de rougissement est bien connu : Herschel lui-même l'a observé, découvrant au télescope que les étoiles faibles étaient toujours rougeâtres, et il l'a même expliqué (déviation préférentielle des rayons bleus par une substance interstellaire), mais il n'en a tenu aucunement compte dans ses calculs stellaires. De plus, depuis le début du $20^e$ siècle, plusieurs astronomes (Hartmann, Frost, Slipher,...) rapportent la présence de raies stationnaires dans les spectres, correctement interprétées comme l'empreinte du milieu interstellaire. Malgré cela, l'extinction est majoritairement jugée négligeable. Cela n'est pas toujours faux : en effet, l'absorption a peu d'importance pour les nébuleuses spirales ou la plupart des amas globulaires, situés hors du plan galactique où se concentre cette matière interstellaire – ce qui explique probablement pourquoi Shapley comme Curtis la négligent sans état d'âme – mais cela s'avère crucial pour l'interprétation des mesures dans la Voie Lactée, et donc les calibrations servant de base à tout le reste. Le tournant se produit en 1930. Trumpler compare alors deux méthodes de mesures de distance affectées de manière différente par l'absorption, et montre que les résultats ne coïncident pas : l'extinction existe bel et bien, et il faut en tenir compte pour obtenir des distances correctes. En plus, cela explique non seulement l'évitement du plan pour les amas globulaires et les spirales, mais aussi l'invisibilité du centre galactique, une objection qu'on avait aussi opposée à Shapley (le centre galactique et son bulbe seront finalement observés lorsque l'astronomie infrarouge verra le jour).

En parallèle, Bertil Lindblad (en 1924) et Jan Oort (en 1927) expliquent les vitesses stellaires observées par la rotation galactique, ce qui nécessite d'excentrer le Soleil, comme le suppose Shapley. D'un autre côté, la présence de bras spiraux dans la Voie Lactée est mise en évidence par l'étude des étoiles chaudes (Baade, Morgan) puis par l'émission radio de l'hydrogène : notre Voie Lactée s'avère donc être une « spirale » elle aussi, renforçant sa similitude avec les objets nébuleux ! En outre, Baade découvre en 1944 que les étoiles de notre voisinage et celles des amas sont différentes : il les nomme respectivement population I et II. Cela implique qu'une hypothèse de Shapley est fausse (les étoiles des amas ne sont pas parfaitement similaires aux voisines). Tout cela, combiné à l'extinction, conduit à une révision de la calibration période-luminosité des Céphéides (Baade 1952), et donc à un changement de l'échelle des distances. Au final, la Voie Lactée s'avère moins grande que ce que Shapley ne pensait (la valeur actuelle est de 100 000 années-lumière), mais quand même bien plus grande que ce que pensait Curtis, et les nébuleuses spirales sont extrêmement lointaines (plus même que ce que Curtis avançait).

Dans ce domaine, l'observation décisive s'avère un cruel revers pour Shapley. En 1925, Hubble rapporte sa découverte de Céphéides dans trois nébuleuses, NGC6822, M33 et M31. En utilisant la propre méthode de Shapley, avec sa calibration, il détermine des distances si grandes pour ces objets qu'il ne peut s'agir que d'autres galaxies, externes à la Voie Lactée. « *Here is the letter that destroyed my Universe* » aurait dramatiquement déclaré Shapley lorsqu'il reçut l'annonce de cette découverte (cf. Payne-Gaposchkin C., 1984, an autobiography and other recollections)... mais il a aussi répondu à Hubble que « *your letter... is the most entertaining piece of litterature I have seen for a long time.* » (Gingerich O., 1988, IAU Symposium 126, p23). Shapley aurait pu se consoler en se disant qu'à une telle distance, ces galaxies doivent être très grandes, comme il le suggérait pour la Voie Lactée[15], et que l'utilisation même par Hubble de la relation période-luminosité validait sa méthode, qui ne sera d'ailleurs plus remise en cause par la suite.

Mais ce n'est pas le clou final. L'observation de vitesses très élevées pour les « nébuleuses spirales » (désormais appelées galaxies) avait été utilisée comme argument dans le débat, par Curtis. Tant avant qu'après le débat, de nombreux chercheurs (Eddington, Shapley, Wirtz, Lundmark) soupçonnent pour les spirales une corrélation entre vitesses (ou plus précisémments *redshifts*, soit les décalages vers le rouge) et distances (ou ses proxys, diamètre ou magnitude des objets, Smith R.W., 2009, Journal for the History of Astronomy, 40, 71). Côté théorique, Willem de Sitter en 1917 puis un jeune théoricien belge, Georges Lemaître, en 1927 découvrent une relation similaire en manipulant les équations de relativité générale. Lemaître calcule cette relation en détails et la compare aux observations disponibles, qui s'avèrent confirmer la chose (ce passage n'a cependant pas été traduit en anglais dans la reproduction de l'article original, cf. Livio M., 2011, Nature, 479, 171 et références s'y trouvant). Deux ans après Lemaître, Hubble et Humason arrivent à la même conclusion observationnelle (mais sans l'interprétation théorique, et en reprenant largement les mesures de vitesse de leurs prédécesseurs) – la relation est depuis connue sous le nom de « loi de Hubble », et est considérée comme l'une des meilleures preuves de l'expansion de l'Univers (même si Hubble lui-même ne cautionnait pas cette interprétation).

Quant à l'uniformité de l'Univers, Hubble semble la retrouver en 1926, grâce à des méthodes de comptage, pour les galaxies (en lieu et place des étoiles). Toutefois, la présence d'amas sera confirmée dans les décennies suivantes (Shapley & Ames, de Vaucouleurs, Shane & Wirtanen, Abell) et des observations plus récentes, des travaux pionniers de Vera Rubin dans les années cinquante aux surveys systématiques de la dernière décennie, montrent cependant la présence de structures à grande échelle, filaments et zones vides entre eux.

En résumé, ni Curtis ni Shapley n'avait parfaitement raison, et « leur » modèle ne survécut donc qu'en partie. Toutefois, leur opposition montre l'intensité des travaux de l'époque, qui s'amplifièrent encore par la suite au vu de la controverse. Leur héritage fut une transformation complète de la construction des cieux. Le débat de 1920 s'avère donc bien une charnière, un climax qui marque la naissance de la cosmologie actuelle. À noter que l'incommunicabilité pour cause de spécialisation n'a fait qu'augmenter depuis, forcément.

6. Conclusion

La méthode scientifique veut qu'une seule observation puisse réfuter une théorie toute entière. En pratique, cependant, les conséquences d'une observation ne sont pas toujours prises en compte ou au contraire sont hypertrophiées. La théorie héliocentrique d'Aristarque fut la conséquence de la déduction d'une taille du Soleil supérieure à celle de la Terre... mais cette constatation resta sans effet réel sur la théorisation de la structure du monde. La théorie copernicienne aurait dû être rejetée au 17$^e$ siècle au profit de la théorie géohéliocentrique, seule compatible avec les observations disponibles, mais cela ne fut pas le cas, l'élégance théorique prenant le dessus sur des observations décisives. Au 18$^e$ siècle, les comptages d'étoiles permirent de déterminer la taille, finie, de notre Galaxie, et cette méthode sera raffinée alors que son inventeur avait montré son inanité. Enfin, il y a un siècle, deux modèles d'univers s'affrontèrent, chacun tirant sa légitimité d'observations a priori incontestables, et la réconciliation des deux, en quelques décennies, se produisit via leur remplacement par un modèle très différent. Dans tous les cas, les relations entre observations et théories s'avèrent bien moins tranchées qu'on pourrait le croire, et l'acceptation ou non d'une observation se révèle très signifiante quant au climat et à l'évolution de la science de l'époque.

---

15 Mais avec la distance trouvée par Hubble, la Voie Lactée restait quand même plus grande que ses voisines galactiques, ce qui ennuyait fort Shapley et ses collègues. Il semblait en effet étonnant que notre Galaxie soit privilégiée de quelque manière que ce soit. Shapley proposera un modèle hybride (la Voie Lactée combinant plusieurs spirales), mais la solution à ce problème viendra finalement simplement de la révision de la calibration de la relation période-luminosité par Baade en 1952.